# Does Quantity Make a Difference?
# The importance of publishing many papers

Peter van den Besselaar [1] & Ulf Sandström [2]

[1] *p.a.a.vanden.besselaar@vu.nl*
VU University Amsterdam, Department of Organization Sciences & Network Institute

[2] *ulf.sandstrom@oru.se*
Royal Institute of Technology, INDEK & Orebro University, Business School, Orebro

**Abstract**
Do highly productive researchers have significantly higher probability to produce top cited papers? Or does the increased productivity in science only result in a sea of irrelevant papers as a perverse effect of competition and the increased use of indicators for research evaluation and accountability focus? We use a Swedish author disambiguated dataset consisting of 48,000 researchers and their WoS-publications during the period of 2008-2011 with citations until 2014 to investigate the relation between productivity and production of highly cited papers. As the analysis shows, quantity does make a difference.

**Introduction**
One astonishing feature of the scientific enterprise is the role of a few extremely prolific researchers (Price, 1963). Thomson Reuters call them *Highly Cited Researchers* and they are listed and recognized per area. Based on another dataset, Scopus publications, Klavans & Boyack (2015) call them "superstars" and use them for large-scale studies of publication behaviour, thereby showing that superstars publishes less in isolated areas (retrieved using a clustering procedure), in dying areas, or in areas without an inherent dynamics. Highly productive and cited researchers tend to look for the new opportunities. Obviously, the highly productive researchers have to be taken into consideration for many reasons, both for science policy and for scholarly understanding of how the science system works.

Within bibliometrics there is a discussion on how to measure and to identify the superstars. Many current papers discuss the correlation between the various indicators developed for performance measurement. One of the stable outcomes is that there is a high correlation between the numbers of papers a researcher has published and the number of citations received (Bosquest & Combes 2013). From that perspective, both indicators tend to measure the same attribute of researchers, as is actually materialized in the introduction of the H-index (Hirsch 2005). Parallel, the discussion about impact has shifted from counting (field normalized) numbers of citations to more qualified types of citations and publications. As the progress of science rests on the huge amount of effort and publications, the number of real discoveries and path breaking new ideas is rather small. This has led to a different focus. Instead of counting publications and citations, the decisive difference is whether a researcher contributes to the small set of very high cited papers. Different thresholds are deployed, from the top 1% highly cited papers to the top 10% highly cited papers or with the CCS method proposed by Schubert & Glänzel (1988). Only when reaching into these select set of papers that qualifies for citations above the x% level one can be considered as really having distinctive result that contributes to scientific progress. Increasingly, performance measures take this selectivity into account, and when calculating overall productivity and impact figures for researchers, papers (productivity) and citations (impact) are weighted differently depending on the impact percentile the paper belongs to (Sandström & Wold 2015).

Of course, the question now comes up what a good publication strategy is – given this way of performance evaluation. Is publishing a lot the best way – or does that generally lead to normal science, with low impact papers? The total number citations received may still be large, but no top papers may have been produced. This is also the underlying idea of emerging movements in favour of 'slow science' like e.g. in the Netherlands; there the 'science in transition' movement (Dijstelbloem & Huisman 2014) was able to convince the Minister of science and the big academic institutions to remove productivity as a criterion from the guidelines for the national research assessment (SEP). The underlying idea is that quality and not quantity should dominate – and that with all the emphasis on publications this has become corrupted.

However, others seem to see this differently. In his important work on scientific creativity, Simonton has extensively argued that (i) having a breakthrough idea is a low probability event that happens by chance, and therefore that (ii) the more often one tries, the higher the probability to have a 'hit' so now and then (Simonton 2004). There are also other contextual factors that may improve the chance for important results, but overall, the number of tries (publications) is the decisive variable. This also explains why Nobel laureates have so many more publications than normal researchers (Zuckerman 1967; Sandström & Van den Besselaar forthcoming). The more often you try (publish) the higher the probability that there is something very new and relevant, and atypical for the scientific community (Uzzi et al. 2013).

This brings us to the question whether there is a strong positive, or a negative relation between overall output (number of publications) and high impact papers. The answer of this question may inform our understanding of knowledge production and scientific creativity, but is also practically relevant for selection processes, and as explained above for research evaluation procedures: is high productivity a good thing, or a perverse effect and detrimental to the progress of science?

**Methods and Data**
In order to investigate this, we use the 74,000 WoS-publications 2008-2011 (with citations until 2014) of all researchers with a Swedish address using the following document types in databases SCI-E, SSCI and A&HCI: articles, letters, proceeding papers and reviews.

For identifying authors and keeping them separate we use a combination of automatic and manual *disambiguation* methods. An algorithm for disambiguating unique individuals was developed by Sandström & Sandström, based on Soler (2007) and Gurney et al. (2012) and was found to proceed fast, although with minor manual cleaning methods. The deployed method takes into account surnames and first-name initials, the words that occur in article headings, and the journals, addresses, references and journal categories used by each researcher. There is also weighting for the normal publication frequency of the various fields.

As indicated the data covers 74,000 articles and 195,000 author shares that have been judged to belong to Swedish universities or other Swedish organisations. In a few cases, articles from people who have worked both in Sweden and in one or more Nordic countries have been kept together, and articles have thus been included even if they came into being outside Sweden (the process of distinguishing names is thus carried out at Nordic level).

All articles by each researcher are ranked, based on received citations and according to the about 260 subject categories as specified in the Web of Science, and the articles are divided into CSS (Characteristic Scores and Scales) classes (0, 1, 2, 3). While measures based on percentile groups (e.g. top1% etc.) are arbitrarily constructed, CSS have some advantages concerning the identification of outstanding citation rates (Glänzel & Schubert 1988). The CSS method is a procedure for truncating a sample (e.g. a subfield citation distribution) according to mean values from the low-end up to the high-end. Every group created using this procedure helps to identify papers that fulfil the requirements for being cited above the respective thresholds. In this paper we will use two levels, level CSS1 and CSS3, which in the former case cover the 20-25 % most cited papers, and in the latter case the about 2-3 % of most cited papers: the "outstandingly cited papers" (Glänzel 2011).

In this paper we will investigate the relation between quality and quantity in several different ways. We proceed in this way, as from a methodological perspective different options are open, without a convincing argument which one would be the better. By using a variety of methods, we avoid to produce results as artefacts of the method deployed.

(i) Firstly, we calculate the probability to have one, two or three and more top cited papers, given the productivity level. We calculate this for the health, i.e. medical sciences (about 15,000 researchers), where we classify these authors in several productivity classes. Class 1 has one publication in the four years period under study, class 2 has two, class 3 has three to four, class 4 has five to eight, class 5 has nine to sixteen, class 6 has seventeen to 31 publications, and finally class 7 covers researchers with 32 or more publications. Publications are integer counted, but citations are field normalized.

(ii) Secondly, we do a simple regression with the total number of (integer counted, IC) publications as the independent variable, and the (also integer counted) number of top cited publications in terms of one of the definitions as discussed above. Also here citations are field normalized. We have here all researchers, without normalizing for field based productivity figures. As the total set of researchers s dominated by life and medical sciences and by natural sciences, and as these groups have comparable average publications and citations, we assume that this does not really influence the results. Under point four below, we introduce a way of taking field differences in productivity into account.

(iii) Thirdly, we do the same analysis as described above, but use fractional instead of full counting. This helps to investigate the effect of different ways of counting on the relations under study.

(iv) Fourthly, we move to the field-normalized (fractional counted) productivity, and calculate the relation between in this way defined productivity and having at least one publication in CSS1 respectively in CSS3. In the last analysis, we can provide an integrated analysis of all researchers across all fields, as we produced field normalize output counts. This is done with a method – Field Adjusted Production (FAP) based on Waring estimations – as initially developed by Glänzel and his colleagues (Glänzel et al. 1984; Braun et al. 1990; Koski et a. 2011) during the 1980s. FAP is further explained and tested in Sandström & Sandström (2009). Basically, the method is used in order to compensate for differences between research areas concerning the normal rate of scholarly production. For this all journals in the Web of Science have been classified according to five categories (applied sciences, natural sciences, health sciences, economic & social sciences, and art & humanities). Categorisation of journals into macro fields is based on Science Metrix classification of research into five major domains. Note, that in some of the following analysis we will refrain from applying the

Waring method, consequently, instead the analysis will be performed per scientific macro fields (for further information, see < http://science-metrix.com/en/classification>).

**Results**

*(i) Does the probability of high-cited papers increase with productivity?*
We calculated the number of top cited papers (CSS3) for each of the seven productivity classes. From this, figure 1 was created. Clearly, the probability increases with productivity, and this is the case for 1, 2 and 3 or more papers in the CSS3 class. In fact, the relation is slightly different for the three criteria. The higher the criterion, the larger the effect is at the high end of the productivity distribution.

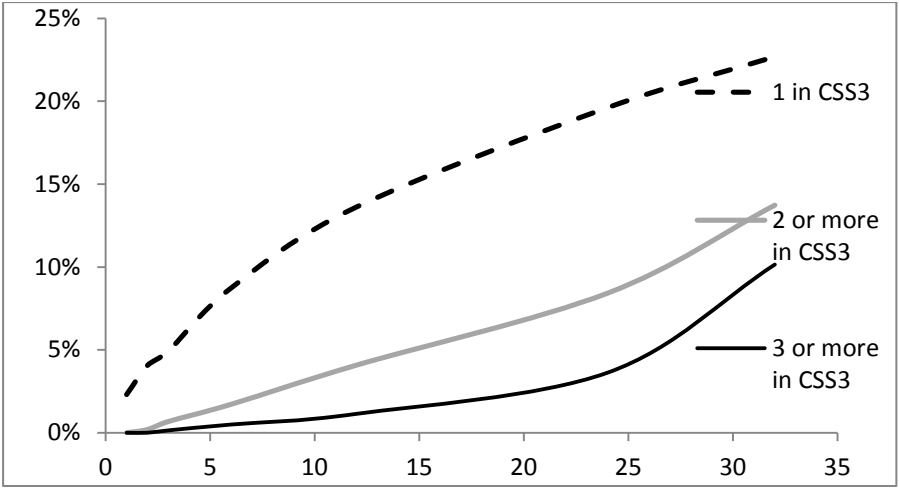
**Fig 1: Share of papers in the CSS3 top cited class by productivity class**

*(ii) What is the effect of productivity on the number of high-cited papers?*
We have done a regression analysis with high-cited papers as dependent variable, and productivity as independent variable. We did the analysis for the various top cited classes. In the three figures below, we show the regression results. For papers in the top 1% cited papers (Figure 2) the correlation is about 0.5. For the CSS3 (Figure 3), the top 10 cited papers (Figure 4), and the CSS1 classes (Figure 5), the correlations are 0.58, 0.78 and 0.88. The correlations are fairly high.

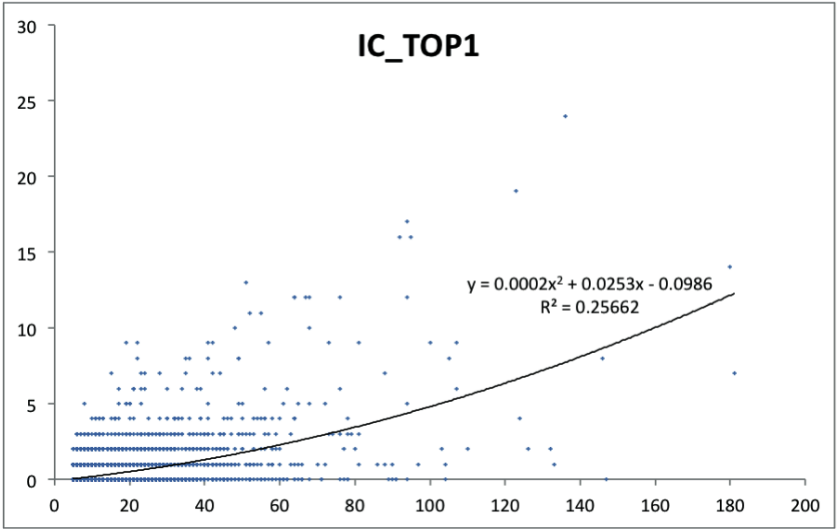
**Fig 2: Top1% cited papers by total number of papers**

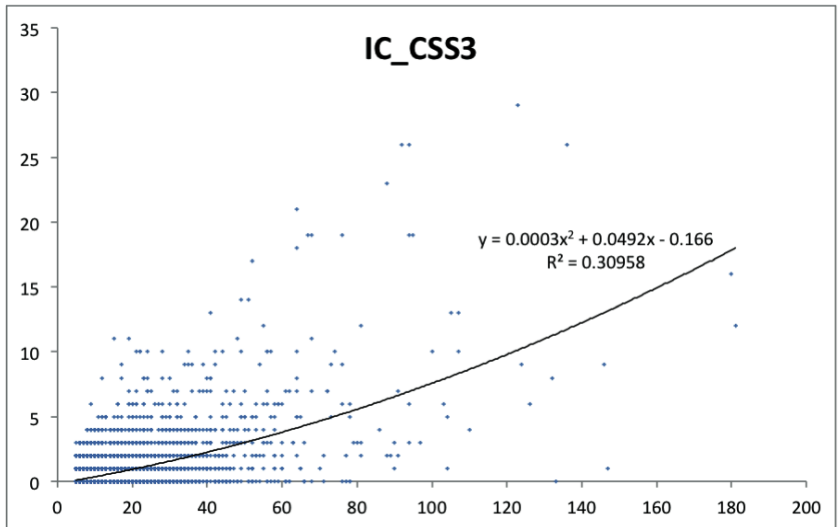
**Fig 3: CSS3 cited papers by total number of papers**

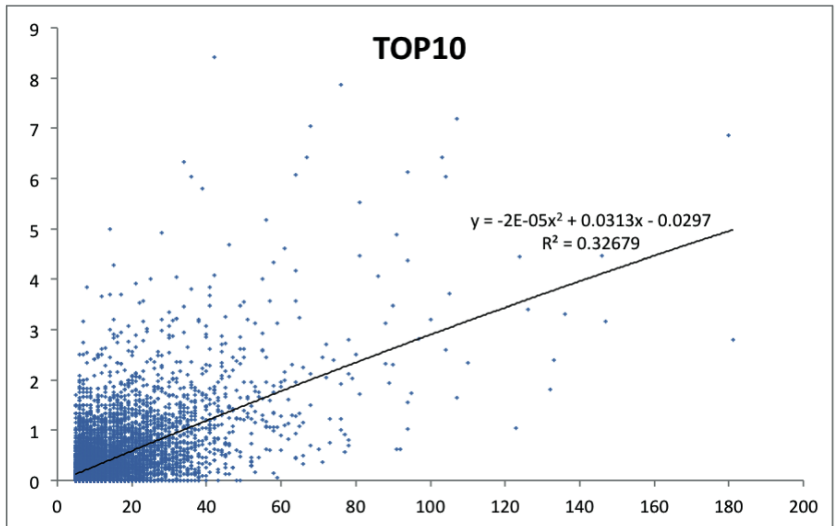
**Fig 4: Top10% cited papers by total number of papers**

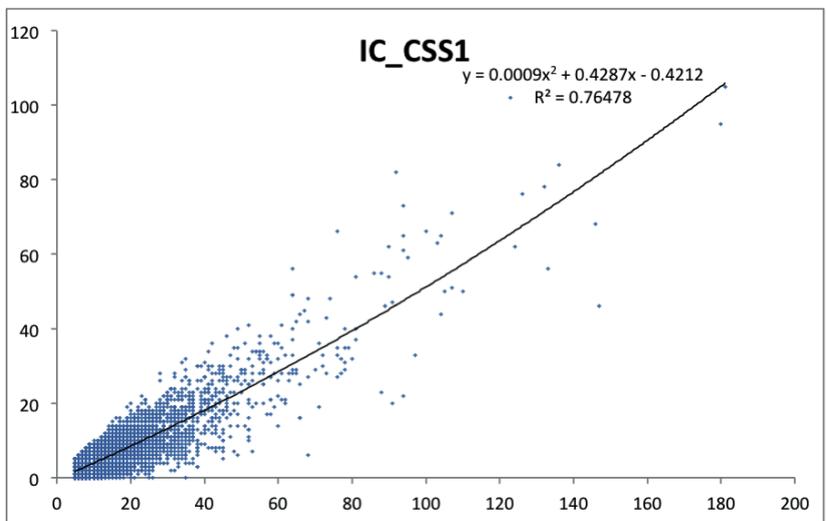
**Fig 5: CSS1 cited papers by total number of papers**

Interestingly, the correlation becomes higher the lower the citation threshold. Why this is the case is not yet investigated. A possibility is that high productive researchers with top papers always have co-authors of these high cited papers who themselves are not highly productive. In that sense one also expects top cited authors in the lower productivity segments, reducing the explained variance. So probably, one should only include PIs in the analysis to avoid this effect. This could be the topic for a subsequent study.

One should realize that a small share of all authors produces most of the papers and of the highly cited papers. The 6.3 % most productive researchers (everybody above eleven publications in four years) are responsible for 37% of all papers and for 53% of the top 1% cited papers. Also this supports the idea that quantity makes a difference.

*(iii) And the effect of fractional counted productivity on the number of high-cited papers?*
We did the above analysis also using fractional counting of productivity. The patterns are the same, but the correlations are about .15 to .20 lower than in the full counted model. How this can be explained will be addressed in a coming paper. But also here, the 6.3% most productive authors are decisive: they have 46.8% of the fractional counted top1% cited papers.

*(iv) What is the effect of field adjusted production counting?*
The relation between having at least one paper in CSS1 and total field normalized output is plotted in Figure 6, and as becomes obvious, the correlation is fairly high (r = 0.79), and not much smaller than in the above where we did not use the field adjusted production (0.90, see Figure 5). The results here suggest that indeed the more papers someone publishes, the higher the probability of having a paper in the group of fairly good papers cited above the threshold of CSS1.

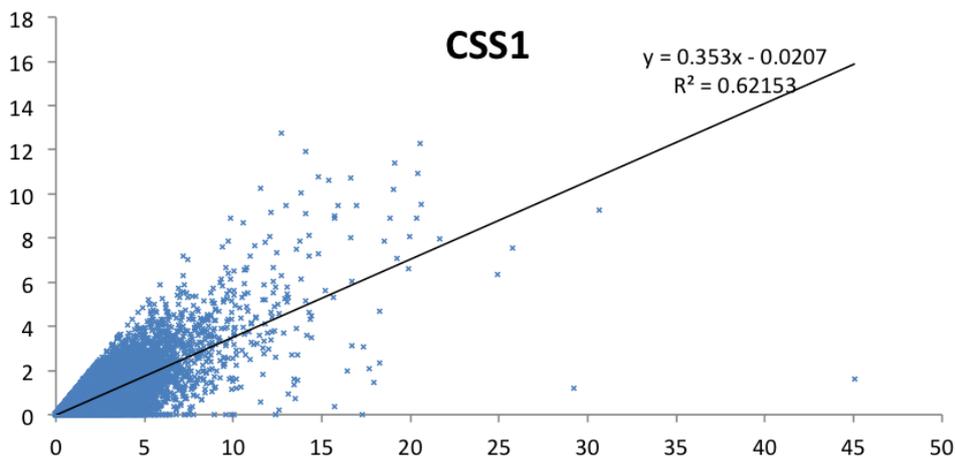

**Figure 6: Fractionalized CSS1 by field adjusted production (all areas of science)**

We also plot the relation between having at least one paper in the CCS3 (Figure 7), so in a much more narrow defined top, and field-normalized productivity, and although correlation is lower here, it is still considerable (r = 0.37). However, in the CSS3 case, the correlation when applying FAP is lower than the correlation without applying FAP (Figure 3), namely is 0.58. These differences need some further exploration.

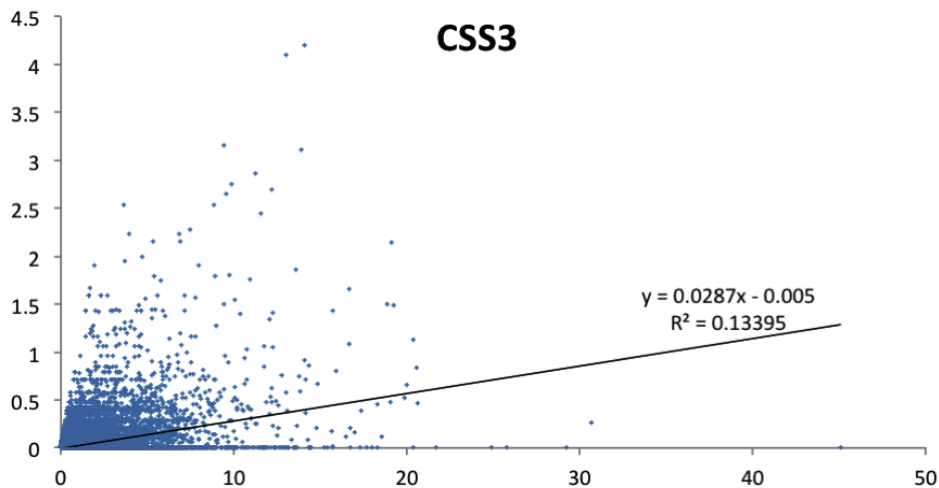

**Figure 7: Fractionalized CSS3 by field adjusted production (FAP - all areas of science)**

The underlying distribution for the fields of Natural sciences and Medical and Life sciences are given in Table 1, which shows for seven distinct productivity categories the percentage of Swedish researchers in that category, the average number of papers published in a four-year period, the average fraction of paper production, and of course the percentage of researchers with at least one paper in CCS3.

**Table 1: CSS3 papers by production levels, Health sciences and Natural sciences**

| Category | Medical and life sciences | | | | Natural sciences | | | |
|---|---|---|---|---|---|---|---|---|
| | researchers | Mean P | Frac P | CSS3 | researchers | Mean P | Frac P | CSS3 |
| 1 (1 paper) | 40.8% | 1 | 0.2 | 0.03 | 9,0% | 1 | 0.2 | 0.02 |
| 2 (2 papers) | 16.92% | 2 | 0.4 | 0.06 | 16,3% | 2 | 0.5 | 0.05 |
| 3 (>2-4) | 17.08% | 3.4 | 0.7 | 0.10 | 17,4% | 3.4 | 0.9 | 0.10 |
| 4 (>4-8) | 13.36% | 6.1 | 1.3 | 0.21 | 13,7% | 6.1 | 1.6 | 0.21 |
| 5 (>8-16) | 7.23% | 11.6 | 2.4 | 0.44 | 8,3% | 11.5 | 2.8 | 0.40 |
| 6 (>16-32) | 3.36% | 22.3 | 4.4 | 1.05 | 4,1% | 22.0 | 4.7 | 0.87 |
| 7 (>32) | 1.18% | 50.5 | 8.8 | 3.45 | 1,2% | 47.6 | 9.8 | 2.68 |
| Average | | 4.3 | 0.9 | 0.17 | | 4.6 | 1.1 | 0.17 |

Data for this table is built on publications from 37,114 researchers.

As 'field adjusted' production (FAP) might be a rather abstract concept, we have translated it below for the various disciplines into 'normal papers'. So, what is the relation between the number of papers produced (in a period of four years) and the probability of having a 'top cited paper' (in the top 2-3% cited papers CSS3 class) during the period 2008-2014? This is a more sophisticated version of the analysis presented in section (i) above. As we clearly see in Table 2, the higher the number of papers, the more likely that one has a paper that ends up to be an outstandingly cited paper. Actually the increase is rather steep and one may say that in most disciplines only with some ten papers in the period under consideration, there is a good chance of having a top paper. The humanities have a different pattern, as with a production of five papers one has the highest chance of reaching the top.

**Table 2: Probability of one outstanding paper (CSS3) at different levels of production.**

| Average # of publications | Class | Discipline | | | | |
|---|---|---|---|---|---|---|
| | | Natural | Health | Applied | Ec &Soc | Hum |
| 1 | 1 | 5% | 7% | 7% | 6% | 9% |
| 2 | 2 | 11% | 13% | 13% | 13% | 8% |
| 3 | 3 | 20% | 21% | 21% | 24% | 25% |
| 6 | 4 | 31% | 34% | 33% | 34% | 33% |
| 11 | 5 | 49% | 54% | 53% | 55% | 33% |
| 20 | 6 / 7 | 61% | 80% | 66% | 83% | |
| 38 | 7 | | | 88% | | |
| 46 | 7 | 83% | | | | |
| 49 | 7 | | 93% | | | |

Note: Data for this table consist of ≈190,000 article shares with <40 authors per paper.
Note: the numbers of publication are the field-specific averages per productivity class
(for more information, see Table 1).

**Conclusions**

As the above results show, there is not only a strong correlation between productivity (number of papers) and impact (number of citations), that also holds for the production of high impact papers: the more papers, the more high impact papers. In that sense, increased productivity of the research system is not a perverse effect of output oriented evaluation systems, but a positive development. It strongly increases the occurrence of breakthroughs and important inventions (c.f. Uzzi et al 2013), as would be expected from a theoretical perspective on scientific creativity (Simonton 2004). Also other recent work points in the same direction (Larivière & Costas 2015). The currently upcoming discussion that we are confusing quality with quantity therefore lacks empirical support. As we deployed a series of methods, with results all pointing in the same direction, the findings are not an artefact of the selected method.

The analysis also gives an indication of the output levels that one may strive at when selecting researchers for grants or jobs. To produce high impact papers, certain output levels seem to be required – of course field depending.

Future work in this research line will cover various extensions: Firstly, we plan to extend the analysis to some other countries, which of course requires large-scale disambiguation of author names. Secondly, we will in a next version control for number of co-authors, and for gender. The former relates to the discussion about team size and excellence, the latter to the ongoing debate on gender bias and gendered differences in productivity. Thirdly, the aim is to concentrate on principle investigators, and remove the incidental co-authors with low numbers of publications, as they may seem to be high impact authors at the lower side of the performance distribution. This all should lead to a better insight in the relation between productivity and impact in the science system.